\documentclass[12pt]{article}
\usepackage{epsfig,epsf}

\def\beq{\begin{equation}}
\def\eeq{\end{equation}}
\def\eq#1{{Eq.~(\ref{#1})}}

\def\npb#1#2#3{    {\it Nucl. Phys. }{\bf B#1} (19#2) #3}
\def\plb#1#2#3{    {\it Phys. Lett. }{\bf B#1} (19#2) #3}
\def\prd#1#2#3{    {\it Phys. Rev. }{\bf D#1} (19#2) #3}

\def\zpc#1#2#3{    {\it Z. Phys. }{\bf C#1} (19#2) #3}

\begin{document}

\title{
   \hfill {\bf\normalsize FERMILAB-PUB-00/035-T }\\
   \hfill {\bf\normalsize BNL-NT-00/14}\\
   \hfill {\bf\normalsize \today}\\[1cm]
{\bf \Large  Soft Double--Diffractive Higgs Production\\ at Hadron Colliders}}

\author{{\large \bf 
Dmitri Kharzeev\thanks{Email:  kharzeev@bnl.gov} ~$\mathbf{{}^{a)}}$
\quad and
\quad  Eugene Levin \thanks{Email:  leving@post.tau.ac.il;
elevin@quark.phy.bnl.gov} ~$\mathbf{{}^{b),c)}}$}\\[3mm]
{\it \normalsize $^{ a)}$ Physics Department and RIKEN-BNL Research Center,}\\
{\it \normalsize Brookhaven National Laboratory,}\\
{\it \normalsize Upton, NY 11973 - 5000, USA}\\[2mm]
{\it\normalsize $^{ b )}$ HEP Department, School of Physics,}\\
{\it \normalsize  Raymond and Beverly Sackler Faculty of Exact Science,}\\
{\it \normalsize  Tel Aviv University, Tel Aviv 69978, ISRAEL}\\[2mm]
{\it \normalsize $^{ c)}$  Theoretical  Physics Department,}\\
{\it \normalsize  Fermi National
Accelerator Laboratory,}\\
{\it \normalsize P.O. Box 500, Batavia, Illinois  60510, USA}}


 \date{}
\maketitle
\thispagestyle{empty}
\begin{abstract}
We evaluate the non--perturbative contribution to the double--diffractive 
production of the Higgs boson, which arises due to the QCD scale anomaly 
if the mass of the Higgs $M_H$ is smaller than the mass of the 
top quark $M_T$, $M_H < M_T$.
The cross section appears
 to be larger than expected from  
perturbative calculations; we find 
$\sigma_H = 0.019 \div 0.14\, {\rm pbarn}$ at the Tevatron energy, and 
$\sigma_H = 0.01 \div 0.27\, {\rm pbarn}$ at the energy of the LHC.   
 
\end{abstract}

\newpage
\section{Introduction}

 In this letter we suggest a new mechanism for ``soft" double--diffractive production 
of Higgs boson. We consider three reactions

\begin{eqnarray}
p + p & \longrightarrow  & p + [LRG] + H + [LRG] + p \,\,;
\label{R1}\\
p + p & \longrightarrow  & X_1 + [LRG] + H + [LRG] + X_2 \,\,;
\label{R2}\\
p + p & \longrightarrow  & p + [LRG] + H + [LRG] + X_2\,\,;\label{R3}
\end{eqnarray} 
where LRG denotes the large rapidity gap between produced particles and
$X$ corresponds to a system of hadrons with masses much smaller than the
total energy. These reactions have such a clean signature for experimental
search (see Fig.1, where the lego - plot is shown for reaction of \eq{R1})
that they have been the subject of continuing theoretical studies during this decade
( see Refs.\cite{DP1,DP2,DP3,DP4,DP5,DP6}).  

\begin{figure} 
\vspace{-0.4cm}
\begin{center}
\epsfig{file=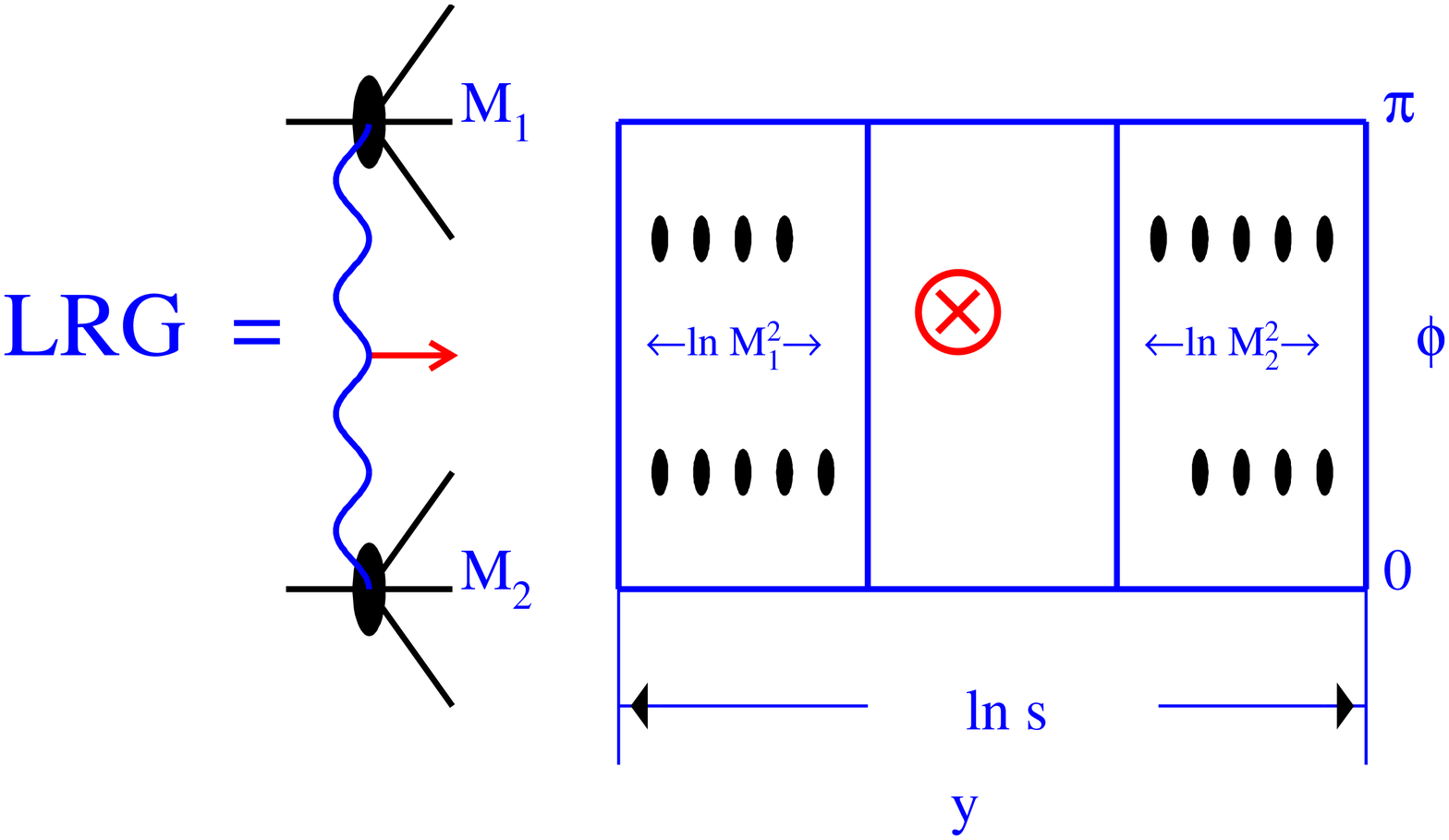,width=15cm,height=8cm}
\end{center}
\caption{\it Lego-plot for double Pomeron Higgs production process.}
\label{lego}
\end{figure}
The main idea behind all calculations, starting from the Bialas-Landshoff paper
\cite{DP1}, is to describe the reactions of \eq{R1} and \eq{R2} 
as a double Pomeron (DP) Higgs production ( see Fig.2 ) .
In Fig.2, the Pomerons are the so--called ``soft" Pomerons for which one uses the
phenomenological Donnachie-Landshoff form ( see Ref. \cite{DL} ),
while the vertex $\gamma$ can be calculated in perturbative QCD.
\begin{figure}
\vspace{-0.4cm}
\begin{center}
\epsfig{file=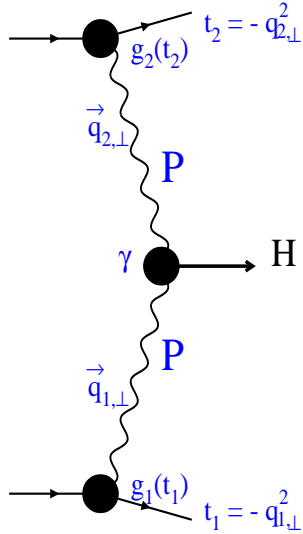,width=5cm,height=8cm}
\end{center}
\caption{\it Double Pomeron Higgs production process.}
\label{fig2}
\end{figure}

We can demonstrate the  problems and uncertainties of such kind of approach
by considering the simplest pQCD diagram  for the double Pomeron
Higgs
production ( DPHP ) (see Fig.3-a). This diagram leads to the amplitude
\beq \label{M1}
M (q q \rightarrow q H q) =
 \frac{2}{9} \,2\,g_H \int
\frac{d^2 Q_{\perp}}{Q^2_{\perp}\,Q^2_{1,\perp}\,Q^2_{2,\perp}}\,4 \,
\alpha_S (Q^2_{\perp})\,( \vec{Q}_{1,\perp}\,\cdot\,\vec{Q}_{2,\perp} )\,,
\eeq
where $g_H$ is the Higgs coupling that 
has been evaluated in perturbative QCD \cite{GH}.
For the reaction of \eq{R1}, $|t_1| = | \vec{Q}_{\perp} - \vec{Q}_{1,\perp}|
\,\approx\,|t_2| =     |
\vec{Q}_{\perp} -
\vec{Q}_{2,\perp} | \,\approx\,2/B_{el}$ and therefore,
\beq \label{M2}
M (q + q \rightarrow q + H + q )\,\propto\,\int\frac{d^2
Q_{\perp}}{Q^4_{\perp}}.
\eeq
\eq{M2} has an infrared divergence which is regularized by the size of the
colliding hadrons. In other words, one can see that already the simplest diagrams
show that the  DP Higgs production is, in a sense, a ``soft" process.
Taking into account the emission of extra gluons denoted in Fig.3-b  as
Pomeron builders, we recover the exchange of the ``soft" Pomerons.

\begin{figure}
\vspace{-0.4cm}
\begin{center}
\begin{tabular}{c c }
\epsfig{file=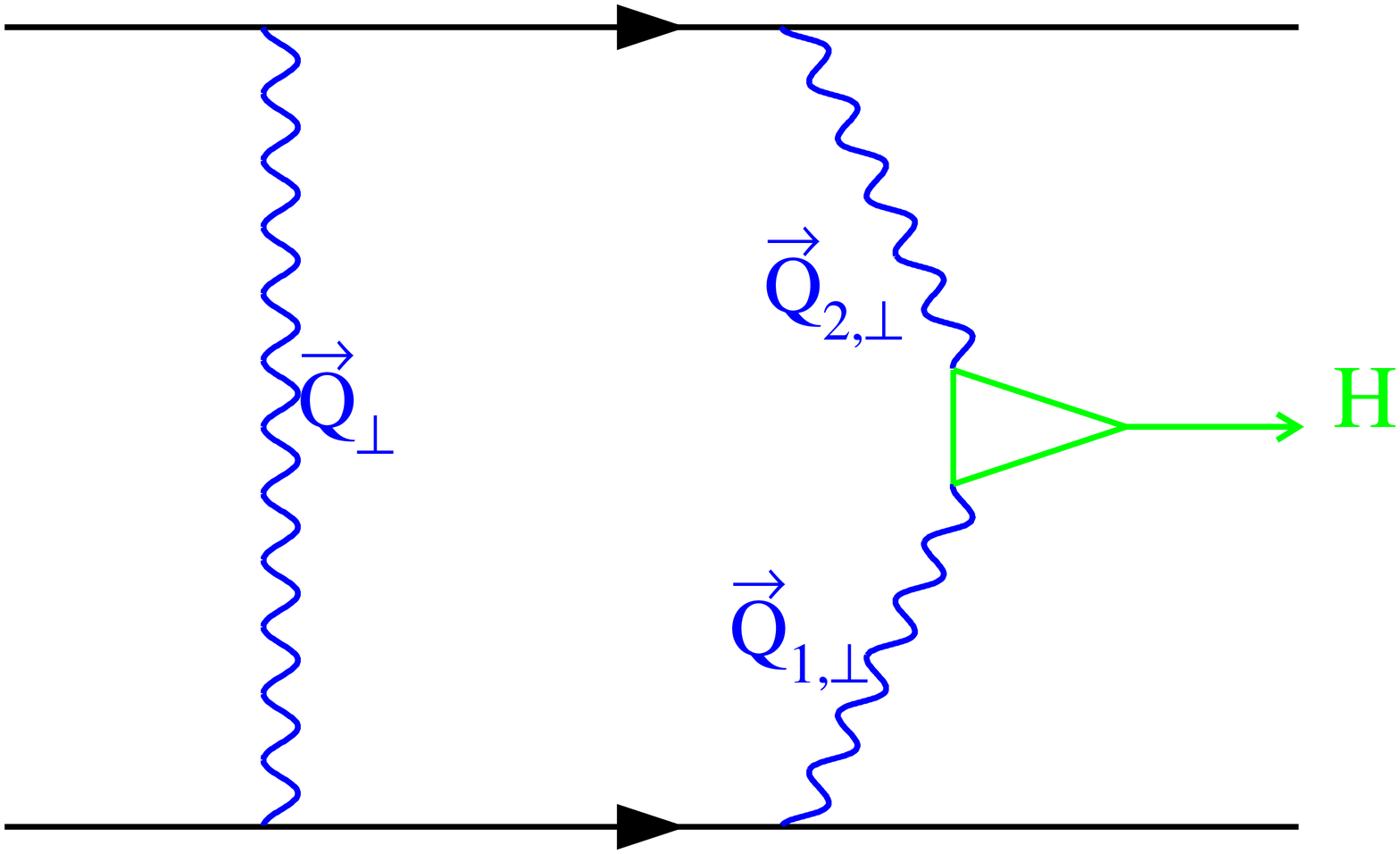,width=8cm}&
\epsfig{file=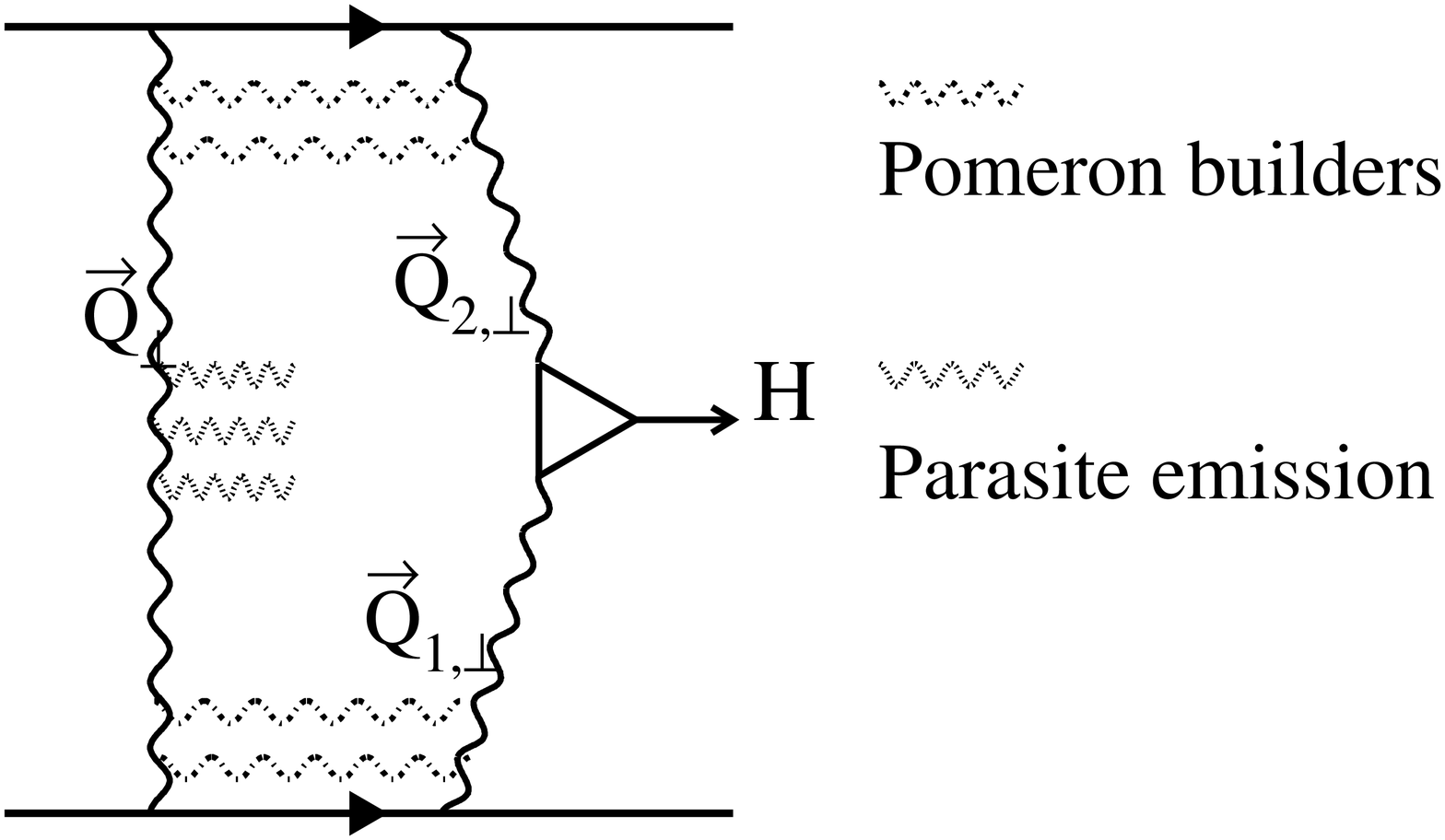,width=8cm}\\
Fig. 3-a & Fig.3-b
\end{tabular}
\end{center}
\caption{\it Double Pomeron Higgs production in the Born approximation
 (Fig.  3-a ) and in leading log approximation (Fig. 3-b ) of pQCD.}
\label{fig3}
\end{figure}

Nevertheless, the emission vertex for the Higgs boson can still 
be calculated in pQCD
since the typical distances inside the quark triangle in fig.3-a are rather
short $\propto 1/M_T$, where $M_T$ is the mass of t-quark.
The coupling $g_H$ has been evaluated in Ref.\cite{GH}
\label{GH} and is given by 
\beq
 g^2_H\,=\, \sqrt{2}\ G_F\ \alpha^2_S(M^2_H)\ N^2/9 \pi^2,
\eeq
where $N$ is a function of the ratio $M_T/M_H$ which was calculated in
Refs.
\cite{GH,DP3}.

In this paper we consider an alternative approach to DPHP, in which we
estimate the value of the cross section from non--perturbative QCD.   
In section 2 we review a non--perturbative method suggested by Shifman,
Vainshtein and Zakharov  \cite{SVZ} for the evaluation of the 
coupling of the Higgs boson to 
hadrons; it is valid if the mass of the Higgs is smaller than the 
mass of the top quark. 
In section 3  we develop a method of obtaining the DPHP cross
section using the approach of Ref. \cite{SVZ}. The problem of survival of
large rapidity gaps (LRG) will be discussed in section 4. We conclude
in section 5 with discussion of our results and of the uncertainties
inherent to our approach.

\section{The coupling of Higgs boson to hadrons in\\ non-perturbative QCD}

To evaluate the non--perturbative coupling of the Higgs boson to hadrons,  
we need to have a closer look at the properties of the energy--momentum 
tensor in QCD. The trace of this tensor is given by
\beq
\Theta^{\alpha}_{\alpha} = {\beta (g) \over 2g} 
G^{\alpha\beta a}G^a_{\alpha\beta} + \sum_{l=u,d,s} m_l (1 + \gamma_{m_l})
\bar{q_l}q_l + 
\sum_{h=c,b,t} m_h (1 + \gamma_{m_h})\bar{Q_h}Q_h, \label{3.7}
\eeq   
where $\gamma_{m}$ are the anomalous dimensions; in the following we
will assume that the current quark masses are redefined as $(1 + \gamma_m)m$. 
The appearance of the scalar gluon operator in (\ref{3.7}) is the consequence 
of scale anomaly \cite{scale}, \cite{scale1}.   
The QCD beta function can be written as 
\beq
\beta (g) = - b {g^3 \over 16\pi^2} + ..., \ b = 9 - {2 \over 3} n_h,
\label{3.8}
\eeq
where $n_h$ is the number of heavy flavors ($c,b,..$). 
Since there is no valence heavy quarks inside light hadrons,  
at scales $Q^2<4m_h^2$ 
one expects decoupling of heavy flavors. 
This decoupling was consistently treated in the framework of the 
heavy-quark expansion \cite{SVZ}; to order $1/m_h$, 
only the triangle graph with 
external 
gluon lines contributes. Explicit calculation shows \cite{SVZ} that 
the heavy-quark 
terms transform in the piece of the anomalous gluonic part of $\Theta^{\alpha}
_{\alpha}$:
\beq
\sum_{h} m_h \bar{Q_h}Q_h \to -{2\over 3}\ n_h\ {g^2\over 32\pi^2}  
G^{\alpha\beta a}G^a_{\alpha\beta} + ... \label{3.9}
\eeq
It is immediate to see from (\ref{3.9}), (\ref{3.7}) and (\ref{3.8}) 
that the heavy-quark 
terms indeed cancel the part of anomalous gluonic term associated 
with heavy flavors, so that the matrix element of the energy--momentum 
tensor can be rewritten in the form
\beq
\Theta^{\alpha}_{\alpha} = {\tilde{\beta} (g) \over 2g} 
G^{\alpha\beta a}G^a_{\alpha\beta} + \sum_{l=u,d,s} m_l
\bar{q_l}q_l, \label{3.10}
\eeq 
where heavy quarks do not appear at all; the beta function in (3.10) 
includes the contributions of light flavors only:
\beq
\tilde{\beta} (g) = - 9 {g^3 \over 16\pi^2} + ...
\label{3.11}
\eeq

Because the mass of the Higgs boson $M_H$ is presumably large, 
its coupling to hadrons involves the knowledge of hadronic 
matrix elements at the scale $Q^2 \sim M_H^2$, at which the heavy 
quarks in general are not expected to decouple. However, 
if the Higgs boson mass $M_H$ is smaller than the mass of the 
top quark $M_T$, one can still perform expansion in the ratio 
$M_H/M_T$; we expect this to be a reasonable procedure if 
$M_H \leq 100$ GeV.
In this case, one finds 
\beq \label{rel1}
M_T \bar{t}t \to -{2\over 3}\  {g^2\over 32\pi^2}  
G^{\alpha\beta a}G^a_{\alpha\beta} + ... 
\eeq 

Since the mass of the hadron is defined as the 
forward matrix element of the energy--momentum tensor, the expression \eq{rel1} 
leads to the following 
Yukawa vertex for the coupling of a Higgs boson to the hadron:
\beq \label{FINHH}
2^{\frac{1}{4}} G^{\frac{1}{2}}_F \cdot H \phi^2_h \,\cdot\,\langle h |
M_T \bar t t | h \rangle \,\,=\,\,2^{\frac{1}{4}} G^{\frac{1}{2}}_F \cdot
H \phi^2_h \,\cdot\,\frac{2 M^2}{27}\,\,;
\eeq
this relation is valid in the chiral limit of massless light quarks (see \eq{3.10}); 
$M_T$ is the mass of the heavy quark and $M$ is the hadron mass.
We put the number of light quarks $N_F = 3$ and the number of
colors $N_c = 3 $; $\phi_h$ and $H$ are hadron and Higgs operators.
Note that, as a consequence of scale anomaly, \eq{FINHH}  
does not have an explicit dependence on the coupling $\alpha_s$.

\section{Estimates for double Pomeron Higgs production cross sections}
 
\subsection{General formulae for double Pomeron Higgs production}
The amplitude for Higgs production in the Pomeron approach is given by
( see for example  Refs.\cite{DP1,GLM1}) 
\beq \label{DP1}
M (h +h  \rightarrow h + H + h
)\,\,=\,\,g_1(t_1)\,\cdot\,g_2(t_2)\,\cdot\,\gamma(t_1,t_2)\,\cdot\,
\eta_{+}(t_2)\eta_{+}(t_1)\,\cdot\,
\left(\,\frac{s}{s_2} \right)^{\alpha_P(t_2)}\,\cdot\,
\left(\,\frac{s}{s_1} \right)^{\alpha_P(t_1)}\,\,,
\eeq
where $s_1 = ( P_1 + q_1 )^2 $ and $s_2 = (P_2 + q_1 )^2$ ( $P_{1,2}$ are
momenta of incoming hadrons ); $\eta_{+}(t_i)$ is a signature factor,
which for the Pomeron is 
\beq \label{DP2}
\eta_{+}(t_i)\,\,\,=\,\,\, i +  tan^{-1}\left( \frac{\pi
\alpha_P(t_i)}{2} \right) \,\,,
\eeq
where $\alpha_P(t)$ is the Pomeron trajectory, $\alpha_P(t) = 1 + \Delta_P +
\alpha'_P\,\,t $, with $\Delta_P \approx  0.08 $ \cite{DL}; 
all other notations are evident from Fig.2.

The cross section for DPHP in the central rapidity region ($y_H = 0 $,
where $y_h$ is the rapidity of the produced Higgs boson) can be written down as  
\beq \label{XS}
\frac{d \sigma}{d y_H d t_1 d t_2}|_{y_H = 0}\,\,=
\eeq
$$\,\,\frac{1}{2 s} | 
M (h +h  \rightarrow h + H + h ) |^2 \prod_{i = 1,2}\,\cdot\, \frac{d^3
P'_i}{(2\pi)^3 2 P'_{i,0}}\,\cdot\frac{ d^2 p_{H,\perp}}{2 (2 \pi)^3}
\,\cdot\,(2 \pi )^4 \,\delta^{(4)}( P_1 + P_2 - P'_1 - P'_2 - p_H)\,\,
$$
where $P'_i$ are momenta of recoil hadrons, while $p_H$ is the momentum of
the produced Higgs boson.
 
Performing all integrations and recalling that $s_1 \cdot s_2 = M^2_H \cdot
s$ we obtain
\beq \label{GF2}
\frac{d \sigma}{d y_H d t_1 d t_2}|_{y_H =
0}\,\,=\,\,\frac{2\,\,g^2_1(t_1)
\cdot g^2_2(t_2) \cdot \gamma^2(t_1,t_2)}{\pi ( 16 \pi )^2}\,\cdot\,\left(
\frac{s}{M^2_H} \right)^{2\,\Delta_P} \,\cdot e^{\alpha'_P\,\ln(
s/M^2_H)\,[ t_1 + t_2 ]}\,\,.
\eeq
 
We will assume that $\gamma(t_1, t_2 ) $ is a smooth function of
$t_1$ and $t_2$ in comparison with $g_1(t_1)$ and $g_2(t_2)$.  Indeed, 
the t-dependence of $g_i$ is related to the quark distribution inside the 
hadron while the t-dependence of $\gamma$ is determined by the mean
transverse of gluon inside the Pomeron. The typical scale for this
momentum is $1/\alpha_P \,\approx\,4 \ GeV^2$ which is much larger than the
typical momentum of a quark in a hadron ( $\approx\,\,0.1 \ GeV^2 $ ).   

Using this assumption together with the simplest Gaussian parameterization
for the vertex $g_i(t_i) = g_i(0) \,\,exp( - R^2_0\,|t_i|)$ we obtain
\beq \label{GF3}
\frac{d \sigma}{d y_H }|_{y_H =0}\,\,\,=\,\,\,\frac{8 g^2_1(0)
\,\,g^2_2(0) }{\pi [16 \pi
B_{el}(s/M^2_H)]^2}\,\cdot\,\gamma^2(t_1=0,t_2 =0 ) \,\cdot\,
\left(\,\frac{s}{M^2_H}\,\right)^{2\,\Delta_P}\,\,.
\eeq
Recalling now the well--known relation between the total and elastic cross sections
for the one Pomeron exchange,
namely,
\beq \label{GF4}
R_{el}(s) \,\,=\,\,\frac{\sigma_{el}(s)}{\sigma_{tot}(s)} \,\,=\,\,\frac{g_1(0)
g_2(0) }{ 16 \pi
B_{el}(s)}\,\,\left(\,\frac{s}{s_0}\,\right)^{\Delta_P}\,\,;
\eeq
where $B_{el} = 4 \,R^2_0 + 2 \alpha'_P\,\ln s$,
one can derive 
\beq \label{GF5}
\frac{d \sigma}{d y_H }|_{y_H
=0}\,\,\,=\,\,\,\gamma^2(t_1=0,t_2=0)\,\times\,\frac{8}{\pi}\,
\times\,R_{el}^2\left(\,\frac{s}{M^2_H} \,s_0 \,\right)\,\,.
\eeq
There is only one unknown factor in  \eq{GF5}, namely,
$\gamma^2(t_1=0,t_2=0)$. In the next subsection we present 
the estimates for this factor using the non-perturbative approach that 
has been discussed in the section 2.
   
\subsection{The production vertex $\gamma(t_1=0,t_2=0)$}

Our  estimate of $\gamma(t_1=0,t_2=0)$ consists of two steps:
\begin{enumerate}
\item \,\,For positive values of $t_1 = t_2 = m^2_{glueball}$ we can
obtain
$\gamma(t_1=m^2_{glueball},t_2=m^2_{glueball})$ from \eq{FINHH};
\item\,\, Using \eq{DP1} we can make the analytic continuation to the
region $t_1 < 0$ and $t_2 < 0 $, which corresponds to the scattering process.
\end{enumerate}
We will assume that there exists a tensor $2^{++}$ glueball which lies on the
Pomeron trajectory, namely, that its mass satisfies the following relation: 
\beq \label{GEQ}
 \alpha_P(t= m^2_{glueball})\,\,\,= \,\,\,1
\,+\,\Delta\,+\,\alpha'_P(0)\,m^2_{glueball}\,\,\,=\,\,\,2\,\,.
\eeq
There is no undisputed experimental evidence for such a meson but lattice
calculations give for its mass $m_{glueball} = 2.4$ GeV \cite{GB}.  
This mass is a little bit higher than can be  expected from \eq{GEQ} with the 
experimental $\alpha'_P(0) = 0.25\,GeV^{-2}$ \cite{DL}. On the other hand,
it is possible to describe experimental data using a smaller value of
$\alpha'_P(0) \approx 0.17\,GeV^{-2}$ which is needed to satisfy \eq{GEQ}
with $m_{glueball} = 2.4\ GeV$, assuming the presence of substantial shadowing
corrections \cite{GLMSOFT}.

\begin{figure}
\vspace{-0.4cm}
\begin{center} 
\epsfig{file=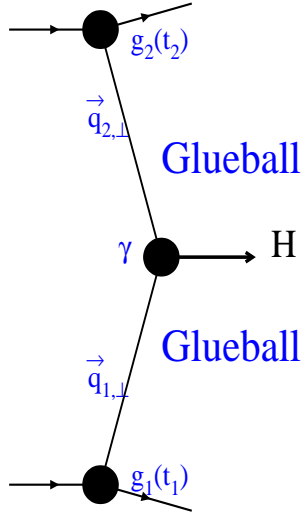,width=5cm,height=8cm}
\end{center}
\caption{\it Higgs emission from glueball.}
\label{fig4}
\end{figure}

For the diagram in Fig.4 the vertex $\gamma_{glueball} $ can be easily evaluated 
from \eq{FINHH}; it is equal to
\beq \label{GA1}
\gamma_{glueball}\,\,\,=\,\,\,2^{\frac{1}{4}}\,
\,G^{\frac{1}{2}}_F\,\,\frac{2\,m^2_{glueball}}{27}\,\,.
\eeq
One can see that \eq{DP1} leads to the contribution described by Fig.4.
Indeed, for $t_i \,\longrightarrow\, m^2_{glueball}$
\beq \label{GA2}
\eta_{+}(t_i)\,\,\longrightarrow\,\,\frac{2}{ \pi\,\,\alpha'_P\,\left(
\,m^2_{glueball}\,\,-\,\,t_i\,\right)}\,\,.
\eeq
(A more detailed discussion of the analytic properties of the Reggeon exchange
can be found in Ref. \cite{REGGE}). 
Using \eq{GA2} and comparing \eq{DP1} with the diagram of Fig.4,  we 
conclude that

\beq \label{GA3}
\gamma(t_1 = m^2_{glueball},t_2 = m^2_{glueball})\,\,\,
=\,\,\frac{\pi}{2}\,\,\alpha'_P(0)\,\,\gamma_{glueball}\,\,.
\eeq

The reggeon approach cannot tell us anything on the relation between
$\gamma(t_1 = m^2_{glueball},t_2 = m^2_{glueball}$ and $\gamma(t_1 = 0,
t_2 = 0 )$.  The only thing that we can claim is that the signature factor takes
into account the steepest part of $t$-behavior.  Therefore, in the next subsection
we will assume that 
\beq \label{GA4}
\gamma(t_1 = m^2_{glueball},t_2 = m^2_{glueball}) \,\,\,=\,\,\gamma(t_1 =
0, t_2 = 0 )\,\,;
\eeq
this is an extreme  assumption which can be used to obtain an upper bound on the 
 cross section.
Uncertainties related to this and other assumptions we make 
will be discussed in detail in section 3.4 
and in the summary, section 5.

\subsection{The magnitude of the cross section}

Using \eq{GA1},\eq{GA3} and \eq{GA4} we can rewrite \eq{GF5} in the simple
form
\beq \label{V1}
\frac{d \sigma}{d y_H}|_{y_H=0}\,\,\,=\,\,\,
2\,\pi\,\left(\alpha'_P\,m^2_{glueball}\right)^2
\times \frac{4\,\sqrt{2}\,\,G_F}{27^2}\times\,R^2\left(
\,\frac{s}{M^2_H} \,s_0 \,\right)\,\,.
\eeq
For $M_H = 100 \ GeV$, the factor $S/M^2_H \cdot s_0 $ is equal to $400 \ GeV^2$ for $s_0
= 1 \ GeV^2$. Therefore, we can take $R_{el} \simeq 0.175$ ( see Fig.5 ) for the
Tevatron energies. 
\eq{V1} leads to
\beq \label{V2}
\frac{d \sigma}{d y_H}|_{y_H=0} (M_h =
100\,\,GeV, \sqrt{s} = 1800\,GeV)\,\,=\,\,6.4\,{\rm pbarn}\,\,,
\eeq
This is a very large number, especially if we recall that the total
inclusive cross section for Higgs production in perturbation theory is on the order of 
 $1$ pbarn
\cite{WM}.  However this estimate does not yet contain the suppression due to the 
(small) probability of the rapidity gap survival, which will be discussed in 
section 4, where we present our final results.
Since $R_{el}$ grows with energy, $R_{el}\,\,\propto\,\,s^{\Delta}$ we expect that 
the cross
section at the LHC energy is approximately 2 times larger than the one in
\eq{V2}.

\begin{figure}
\vspace{-0.4cm}
\begin{center}  
\epsfig{file=  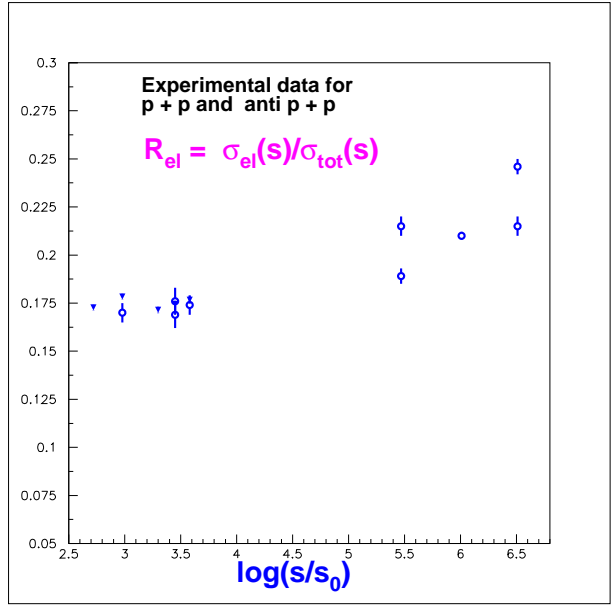,width=8cm}   
\end{center}
\caption{\it Experimental data for the ratio $R_{el}(s) =
\sigma_{el}(s)/\sigma_{tot}(s)$. $s_0$ = $1\,GeV^2$ and the log is taken on
base 10.}
\label{fig5}
\end{figure}

\subsection{Uncertainties of our estimates}

~

1. Let us start with the value of $R_{el}$. We took it from the
experimental data, but we nevertheless have two uncertainties associated with it. 
First,  \eq{GF4} is written for one Pomeron exchange while in experimental
data at $\sqrt{s} \approx \,20\,GeV$ we have about 30\%  contamination
from the secondary Reggeons \cite{DL}. If we try to extract the one Pomeron exchange
from the data, it reduces the value of cross section for DPHP by 1.7
times. Therefore, the value for the cross section can be about $3.8 \ pbarn$
rather than \eq{V2}.   The second uncertainty in evaluation of $R_{el}$ is
the value of $s_0$; even though $s_0 = 1 \,GeV$ appears in all phenomenological
approaches \cite{DL,GLMSOFT},  we have no theoretical argument for the
value of $s_0$. However, since the ratio $R_{el}$ in Fig.5 is a rather smooth
function of energy we do not expect that the uncertainty in the value of $s_0$ 
can introduce a large error.

2.  We can take into account also the reactions of \eq{R2} and \eq{R3}. In \eq{V1}
we would then have to substitute 
\beq \label{V3}
R_{el}\,\,\longrightarrow\,\,R_D = R_{el} +
\frac{\sigma^{DD}(s)}{\sigma_{tot}}\,\,,
\eeq
where $\sigma^{DD}$ is the cross section of the double diffraction
dissociation. Unfortunately, we do not have conclusive data on this cross
section. However, recent CDF  measurements \cite{DDCDF} show that
this cross section could be rather large ( about 4.7 mb at the Tevatron
energy).

3. The principle uncertainty, however, is associated with the continuation
from $t=m^2_{glueball} $ to $t=0$.  This is a question which at present can
only be addressed in the framework of different models. For example, in
Veneziano model \cite{VEN} instead of $\eta_{+}(t) $ (see  \eq{DP2} ) a
new
factor appears, namely
\beq \label{V4} 
\eta^V_{+}(t_i) = \Gamma(2 - \alpha_P(t_i))\,e^{i \frac{\pi
\alpha_P(t_i)}{2}}\,\,.
\eeq
\eq{V4} does not give the factor of $\pi/2$ in \eq{GA3} and, therefore, decreases
the value of the cross section given by \eq{V1} by a factor of 2.5.
We will return to the discussion of the analytic continuation in the summary section.

4. As we have discussed in section 2, we can evaluate  the value of
$\gamma(t_1 = m^2_{glueball},t_2 = m^2_{glueball}) = \gamma_{glueball}$
only if $M_H/M_T < 1$.
 The accuracy of \eq{GA1} is $O( M^2_H/M^2_T)$  and we thus believe 
that \eq{GA1} gives a reasonable estimate of $\gamma_{glueball}$ for
Higgs meson with $M_H \leq 100\,GeV$.

\section{Survival of large rapidity gaps}

 As has been discussed intensively during the past decade (see
Refs.\cite{D1,D2,D3,SPB1,SPB2,SPS1,SPS2,SPS3,SPS4,SPS5,SPS6}), the
cross section of \eq{V1} has to be multiplied 
by a factor $S^2_{spect}$, which is the survival probability of large 
rapidity gap (LRG) 
processes. The ``experimental'' cross section is therefore given by 
\beq \label{SP1}
{\frac {d \sigma (pp \rightarrow pp H)}{d y}}|_{y =0}\,\, =\,\, S^2_{spect}
\frac{d \sigma_P (pp \rightarrow pp H)}{d y}|_{y =0}\,\,.
\eeq
Here, ${d \sigma_P (pp \rightarrow pp H)}/ {d y}$ denotes the cross
section calculated in \eq{V1}. The factor $S^2_{spect}$ has a very
simple meaning -- it is a probability of the absence of inelastic
interactions of the spectators which could produce hadrons
inside the LRG. We have rather poor theoretical control of the value of the
survival probability; this fact reflects the lack of 
knowledge of the ``soft" physics stemming from non-perturbative QCD.
Different models exist, leading to the values about $S^2_{spect} \,\,\approx\,\,10^{-1}$ at the
Tevatron energies. For double Pomeron processes, this quantity has been discussed in 
Ref \cite{GLMDP}. The result of this
analysis is that the value of the survival probability for double Pomeron
production is almost the same as for ``hard" dijet production with LRG
between them. Fortunately, the value of $S^2_{spect}$ has been measured
\cite{LRGD0}, and is equal to 0.07 for the highest Tevatron energy.

Multiplying \eq{V2} by $S^2_{spect}$  = 0.07 and taking into account
suppression due to the factor of \eq{V4}, we obtain 
\beq \label{V5}
\frac{d \sigma}{d y_H}|_{y_H=0} (M_h =
100\,\,GeV, \sqrt{s} = 1800\,GeV)\,\,=\,\,0.2\,{\rm pbarn}.
\eeq
This estimate is not our final result yet, since we still have to correct 
it by the additional suppression factor  $S^2_{par}$ which describes the 
probability of the absence of the parasite gluon emission around the Higgs 
production vertex (see Fig.4-b) \cite{DP6}.   
As was argued in Ref.\cite{DP6},  
\beq \label{SP2}
S^2_{par}\,\,=\,\,e^{ - < N_G(\Delta y = \ln(M^2_H/s_0)) >}
\,\,,
\eeq
with 
\beq \label{SP3}
<N_G(\Delta y = \ln(M^2_H/s_0)) >\,\,\,=\,\,\,\frac{N_{hadrons} (\Delta y
= \ln(M^2_H/s_0))}{N_{hadrons}( one \,\,\,minijet)}\,\,\approx
\,\,\,2\,\div\,4\,\,.
\eeq
It gives $S^2_{par} = 0.14 \,\div\,0.014$.    

The appearance of this factor can be illustrated by the following argument: 
one of the most
important differences between the diagrams of Fig.2 and in Fig.4 is the fact
that the Pomeron exchange is almost purely imaginary while the glueball
exchange leads to the real amplitude. Imaginary amplitude describes the
production of particles and the Pomeron is associated with the inelastic
process with large multiplicity. Therefore, normally, in a large rapidity
region $\Delta y = \ln(M^2_H/s_0)$ we expect to see a large number of produced
particles while in Fig.2 we require that only one Higgs boson is produced. 
Therefore, it seems reasonable to expect a suppression for the double--diffractive 
Higgs production, and this suppression can be described by \eq{SP2} and \eq{SP3}. 

Finally, for the Tevatron energy we expect
\begin{eqnarray} 
\frac{ d \sigma (pp \rightarrow pp H)}{d y}|_{y
=0}\left(\sqrt{s}\,=\,\,1.8\,TeV \right)   \,\,& =&\,\, S^2_{spect}
\,\times\,S^2_{par}\,\times\,\frac{d \sigma_P (pp \rightarrow pp H)}{d
y}|_{y =0}  \nonumber\\
\,\,&=&\,\, \left( \,0.0038 \,\div\, 
0.028\,\right) \,{\rm pbarn}  \label{V6}
\end{eqnarray}

Extrapolating to the LHC energy, we have two effects that work in different directions:
the rise of the Pomeron contribution  and the decrease of the
$S^2_{spect}$ with energy. From Ref. \cite{GLMDP} we expect that 
$S^2_{spect}(\sqrt{s}\,\, =\,\, 16\, TeV )/S^2_{spect}(\sqrt{s}\,\, =\,\,
1.8\, TeV )\,\,\approx\,\,0.7$ while the rise of the Pomeron exchange
leads to an extra factor of 2 in \eq{V6}. Therefore, our final estimate for the LHC is
\begin{eqnarray}
\frac {d \sigma (pp \rightarrow pp H)}{d y}|_{y =0}\left(\,\sqrt{s}
\,=\,16\,TeV\,\right)\,\,& =&\,\, S^2_{spect}
\,\times\,S^2_{par}\,\times\,\frac{d \sigma_P (pp \rightarrow pp H)}{d
y}|_{y =0}\,\nonumber \\ 
&=&\,\,\left( \,0.0015 \,\div\,
0.042\,\right) \, {\rm pbarn}\,\,. \label{V7} 
 \end{eqnarray}
  \eq{V6} and \eq{V7} give significantly larger (by about $5 $ times) larger
cross
sections than expected for double--diffractive production 
in pQCD \cite{KMR}. However, Ref. \cite{DP6} contains an 
estimate of the upper bound on double Pomeron Higgs production in pQCD obtained 
by choosing the largest possible value for $S^2_{par}$ ( see
\eq{SP3} ). This upper bound appears to be about 7 times larger than the 
highest value in \eq{V6}.

\section{Summary and discussion}

The approach suggested in this paper is based entirely on non-perturbative QCD. 
We believe that
such an approach is logically justified for diffractive Higgs production
 since even pQCD calculations
show that this is, to large extent, a ``soft" process (see \eq{M2} and the 
following discussion). 
However, just because of this, we have to stress again that the accuracy of our calculation is not 
very good. We feel, however, that our results 
support the idea \cite{DP6} that in pQCD approach to
 diffractive Higgs production the running QCD coupling has to be taken at 
the ``soft" scale $Q^2 \sim 1\ {\rm GeV}^2 $.  As was argued in
Ref.\cite{DP6}, 
in BLM prescription \cite{BLM} of taking into account the running QCD coupling
one can insert the quark bubbles only in the $t$-channel gluon lines in Fig.
3. Therefore, the running QCD coupling depends on the transverse
momenta of these gluons, and they are determined by the ``soft" scale\footnote{
In this soft regime, the dependence on the coupling constant in the Pomeron 
can disappear as a consequence of scale anomaly \cite{KL}.}. 
The \eq{FINHH} indeed does not depend on the QCD coupling, demonstrating 
the non-perturbative, ``soft" character of the discussed process.

We obtain quite large values for the cross section of the
diffractive Higgs production -- after integration over the Higgs
rapidity $y$ in \eq{V6} and \eq{V7} we get
\beq \label{C1}
 \sigma (pp \rightarrow pp H)\left(\sqrt{s}\,=\,\,1.8\,TeV \right)   \,\,=
\,\,0.019 \,\div\,
0.14\,\,\,\, \,{\rm pbarn}  \,\,.
\eeq
and 
\beq \label{C1pr}
 \sigma (pp \rightarrow pp H)\left(\sqrt{s}\,=\,\,16\,TeV \right)   \,\,=
\,\,0.01 \,\div\,
0.27\,\,\,\, \,{\rm pbarn}  \,\,.
\eeq

Comparing our estimates with the ones based on pQCD \cite{DP1,DP2,DP3,DP4,DP5,DP6}
 we conclude that the lowest of our values of the cross section 
 of double Pomeron Higgs production is about the same as the highest
one in the pQCD approach. However, both  our approach and the pQCD one are
suffering from large uncertainties, stemming from the analytical 
continuation in our approach and from the survival probability of rapidity gap and 
the absence of ``parasite emission''
$S^2_{par}$ in pQCD.  
  
Let us point out that \eq{C1} shows that the double Pomeron Higgs production 
constitutes a substantial part
 of the
total inclusive Higgs production.  Moreover, our calculations lead to an 
additional contribution to the inclusive
 cross section which is
shown in Fig.6.
\begin{figure}
\vspace{-0.4cm}  
\begin{center}
\epsfig{file=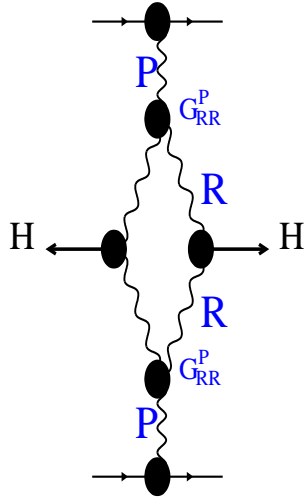,width=5cm,height=8cm}
\end{center}
\caption{\it Mueller diagram \protect\cite{MD} for ``soft" inclusive Higgs
production.}
\label{fig6}
\end{figure}
(Note that the triple Pomeron interaction gives a very small contribution to the process in Fig.
6 due to the small real part in the Pomeron exchange \cite{GLMVF}.)
Using the same approach as in derivation of \eq{GF2} we obtain 
\beq \label{C2}
\frac{d \sigma_{incl}( pp \longrightarrow H + X )}{d y}|_{y_H = 0
}\,\,\,\,=\,\,
\,\,\gamma^2_R (t_1 = 0, t_2 = 0 )\,\,
\frac{ 2 g_1(0) g_2(0) (G^P_{RR})^2 }{\pi (
16
\pi B_R)^2} \,\,\,\frac{
1}{\Delta^2_R}\,\,\left(\,\frac{s}{M^2_H}\,\right)^{\Delta_P}\,\,,
\eeq
where  $\Delta_R \approx 0.5 $.
As a first approximation we can take ( see \eq{GA3} ) 
\beq \label{C3}
\gamma_R (t_1 = 0, t_2 = 0 )\,\,\,=\,\,\,\frac{\pi}{2}\,2^{\frac{1}{4}}
G^{\frac{1}{2}} \,\,\frac{2 \alpha'_R\,m^2_f}{27}\,\,
\eeq
where $m_f$ is the mass of the $f$ - meson which is the first resonance on the secondary
Reggeon trajectory, and $\alpha'_R\,m^2_f = 1.5$.  
Substituting \eq{C3}  in \eq{C2} we obtain 
\beq \label{C4}
\frac{d \sigma_{incl}( pp \longrightarrow H + X )}{d y}|_{y_H = 0 }\,\,=\,\
 \frac{ ( G^P_{RR})^2}{8
\,B_R}\,\times\,\frac{B_R}{B_{el}}\,\times\,2^{\frac{1}{2}}\,G_F\,\frac{1}{3}\,\times\,
R\left( \,\frac{s}{M^2_H}\,s_0 \,\right)\,\,.
\eeq
\eq{C4} gives 
\beq \label{C5}
\frac{d \sigma_{incl}( pp \longrightarrow H + X )}{d y}|_{y_H = 0 }\,\,=\,\
\,\,\left( \frac{G^P_{RR}}{g} \right)^2 \,\,\,3.4\,10^{-7}\,mb\,\,
\eeq
which does not yet contain the suppression arising from the analytical continuation. 
We take this
suppression into account by 
multiplying \eq{C5} by factor $S^2_{par} = 0.14 - 0.014$. 
Unfortunately, we do not know the value for the ratio $G^P_{RR} / g$.
In the triple Pomeron parameterization of the cross section of diffractive
dissociation in hadron reactions \cite{TRR} this ratio changes from 1 to 0.
For $G^P_{RR} / g = 1 $ we get for the ``soft" inclusive cross
section the value of $ 43 \div 430 \,{\rm pbarn} $. On the other hand,  
taking Field and Fox value \cite{TRR} for this ratio
we obtain a much smaller, but still very sizeable  
value of $ 0.43\, \div\, 4.28 \,{\rm pbarn}$. It is thus clear that  
the evaluation of the ``soft'' contribution to the inclusive Higgs production is plagued 
by large uncertainties; however, it might be bigger than the pQCD one \cite{WM}.

\vskip0.3cm

We hope that this paper will help to
look at diffractive Higgs production from a different viewpoint, and will  
stimulate a much needed further work.  To our surprise, despite the very different 
 non--perturbative method used here,  our estimates for the double--diffractive 
production turn out to be not that far from 
the pQCD
calculation \cite{KMR} ( the average is about $5 $ times larger ). It adds some 
confidence in both approaches and gives us a hope that 
one will be able to
perform a reliable calculation in the nearest future.  

\vskip0.3cm

{\bf Acknowledgments:} We are very grateful to Mike Albrow, Andrew Brandt, 
Al Mueller and all
participants of the working group ``Diffraction physics and color coherence"
 at QCD and
Weak Boson Physics Workshop in preparation for Run II at the Fermilab Tevatron
for stimulating discussions of Higgs production and  encouraging  criticism.
We thank Stan Brodsky, Asher Gotsman, Valery Khoze, Larry McLerran, Uri Maor and
Misha Ryskin for fruitful  
discussions on the subject. E.L. thanks the Fermilab  theory department
for creative atmosphere and hospitality during his stay when this paper  
was being written. 

 The work of D.K. was supported by the US Department of Energy 
(Contract \# DE-AC02-98CH10886) and 
RIKEN. 
The research of E.L. was supported in part by the Israel Science 
Foundation, founded by the Israeli Academy of Science and Humanities, 
and BSF \# 9800276.

\end{document}